\documentclass[10pt,twocolumn,english]{revtex4-1}
\usepackage{amsthm}
\usepackage{dsfont}
\usepackage[T1]{fontenc}
\usepackage[latin9]{inputenc}
\usepackage{geometry}
\usepackage{amsmath}
\usepackage{amssymb}
\usepackage{graphicx}
\usepackage{color}

\makeatletter

\usepackage{babel}

\makeatother

\begin{document}

\title{
 Certifying the Potential Energy Landscape}

\author{Dhagash Mehta$^{a}$, Jonathan D.~Hauenstein$^{b}$, David J.~Wales${^{c}}$\\
 \mbox{}\\
$^{a}${\em Dept of Physics, Syracuse University, Syracuse, NY 13244, USA.}\\
$^{b}${\em Dept of Mathematics, North Carolina State University, Raleigh, NC 27695, USA.}\\
$^{c}${\em Dept of Chemistry, The University of Cambridge, Cambridge, CB2 1EW, UK.}}

\begin{abstract} \noindent It is highly desirable for a numerical approximation of a
stationary point for a potential energy landscape
to lie in the quadratic convergence basin of that stationary point.  However,
it is possible that an approximation may lie only in the linear
convergence basin, or even in a chaotic region, and hence not converge to
the actual stationary point when further optimization is attempted.
Proving that a numerical approximation will quadratically converge to the associated
stationary point is termed \textit{certifying} the numerical approximation.
We employ Smale's $\alpha$-theory to stationary points, providing a certification
that serves as a \textit{mathematical proof} that the numerical approximation does indeed
correspond to an actual stationary point, independent of the precision employed.
As a practical example, employing recently developed certification algorithms,
we show how the $\alpha$-theory can be used to
certify all the known minima and transition states of Lennard-Jones LJ$_{N}$
atomic clusters for $N=7,\dots,14$.
\end{abstract}

\maketitle

\noindent\textbf{Introduction:} The surface defined by a potential,
$V({\bf x})$, with ${\bf x}=(x_{1},\dots,x_{n})$,
is called the potential energy landscape (PEL) of the corresponding
physical or chemical system \cite{Wales:04,RevModPhys.80.167}. The
critical points of a PEL, defined by the solutions of the equations
$\partial V({\bf x})/\partial x_{i}=0$ for $i=1,\dots,n$, provide important
information about the landscape.  These critical points, the
stationary points (SPs) of the PEL, can be classified according to
the number of negative eigenvalues of the Hessian matrix,
$H_{i,j}=\partial^{2}V(x)/\partial x_{i}\partial x_{j}$,
evaluated at the SPs: the SPs with no negative eigenvalues are
minima, and the SPs with exactly $I$ negative eigenvalues are called
saddles of index $I$. SPs at which $H$ has one
or more zero eigenvalues in addition to those determined by translational
and rotational symmetry are called singular SPs, or non-Morse points.

Except for rare examples, it is not usually possible to obtain the
SPs analytically, and one has to rely upon computing
numerical approximations by solving the corresponding equations.
For a numerical approach, ``solve'' means
``to compute a numerical approximation of the associated solutions.''
Once a numerical approximation of a solution is obtained,
it is heuristically validated.  Two standard approaches
are to monitor iterations of Newton's method and to substitute the approximations
into the equations to see if they are satisfied up to a chosen tolerance.
Although such a validation usually works well in practice, it does not guarantee that the
numerical approximation will indeed converge quadratically to the
associated solutions using arbitrary precision. In other words, even
if a numerical approximation is heuristically validated, it could
correspond to a nonsolution when using higher precision.
Additionally, Newton iterations may have unpredictable behavior, such as
attracting cycles and chaos, when applied to points that are
not in a basin of attraction \cite{mezey81b,mezey87,wales92,wales93d} of some solution.

If the given system is a set of polynomial
equations, then one can use numerical polynomial homotopy continuation
\cite{Mehta:2009,Mehta:2009zv,Mehta:2011xs,Mehta:2011wj,Kastner:2011zz,Maniatis:2012ex,Mehta:2012wk,Hughes:2012hg,Mehta:2012qr,He:2013yk,SW:05}
to compute all the isolated solutions.
Due to the numerical computations used with this method,
one obtains numerical approximations of the isolated solutions and
hence the aforementioned difficulties also arise.

A proper validation of a numerical approximation is termed \textit{certification},
i.e., a verification that the given numerical approximation will converge quadratically
to the nearby associated solution using
arbitrary precision.  Roughly speaking, quadratic convergence doubles the number
of correct digits after each iteration, so that the associated solution
can be approximated to a given accuracy efficiently.
Starting in the 1980's, Smale and others developed a
method that certifies a numerical approximation as an actual solution of the
system in the following way \cite{BCSS}.
For a given system of equations $f = 0$
and a given point $x^*$, one computes a number $\alpha(f,x^*)$ which,
if it less than $\left(13-3\sqrt{17}\right)/4\thickapprox0.157671$,
guarantees that Newton's method starting from $x^{*}$ will quadratically converge to a
solution of $f = 0$.  Moreover, by using such a certification scheme,
we ensure that our numerical approximations of solutions
are good enough so that more accurate approximations of the solutions
can be obtained easily and efficiently.

\noindent\textbf{Smale's $\alpha$-Theory: } We
summarize Smale's $\alpha$-theory following Ref. \cite{2010arXiv1011.1091H},
where we restrict ourselves to systems of equations that have the same
number of equations as variables, termed square systems.
We should also emphasize that Smale's $\alpha$-theory
is usually used to certify complex solutions for systems of polynomial equations,
so we start with the key points of the theory for this case \cite{BCSS}.
Since more can be said about real solutions, we will
discuss them separately below, as well as generalizations to other types of nonlinear equations,
such as the those involving exponentials and trigonometric functions.

For a system $f$ of $n$ multivariate polynomial equations in $n$ variables,
we denote the set of solutions of $f = 0$
as $\mathcal{V}(f):=\{{\bf z}\in\mathbb{C}^{n}|f({\bf z})=0\}$
and the Jacobian of $f$ at ${\bf x}$ as $J_{f}({\bf x})$.
Consider the Newton iteration of $f$ starting at ${\bf x}$ defined by
$$
N_{f}({\bf x}):=\begin{cases}
{\bf x}-J_{f}({\bf x})^{-1}f({\bf x}), & \mbox{if }J_{f}({\bf x})\mbox{ is invertible,}\\
{\bf x} & \mbox{otherwise.}
\end{cases}
$$
For $k\geq1$, the $k$-th Newton iteration is simply
$$
N_{f}^{k}({\bf x}):=\underbrace{N_{f}\circ\dots\circ N_{f}}_{\hbox{$k$ times}}({\bf x}).
$$

A point ${\bf x}\in\mathbb{C}^{n}$ is called an {\em approximate solution}
of $f$ with {\em associated solution} ${\bf z}\in\mathcal{V}(f)$ if, for each $k\geq1$,
{\small
$$
\left\|N_{f}^{k}({\bf x})-{\bf z}\right\|\leq\left(\frac{1}{2}\right)^{2^{k}-1}\left\|{\bf x}-{\bf z}\right\|,
$$}
where $\|\cdot\|$ is the standard Euclidean norm on $\mathbb{C}^{n}$.
In other words, ${\bf x}$ is an approximate solution to $f$ if it is in the
quadratic convergence basin defined by Newton's method of some solution ${\bf z}$.
The key to Smale's $\alpha$-theory, as shown in the following theorem, is a
sufficient condition for proving that a given point is an
approximate solution without knowledge about ${\bf z}$.

\textbf{Theorem: } If $\alpha(f,{\bf x})<\left(13-3\sqrt{17}\right)/4$
for a square polynomial system $f$ and point ${\bf x}$, then ${\bf x}$ is an approximate solution to $f$ where
{\small
$$\begin{array}{cl}
\alpha(f,{\bf x}):= & \beta(f,{\bf x}) \gamma(f,{\bf x}),\\
\noalign{\smallskip}
\beta(f,{\bf x}):= & \|J_{f}({\bf x})^{-1} f({\bf x})\|, \quad\ \mbox{and}\\
\noalign{\smallskip}
\gamma(f,{\bf x}):= & \underset{k\geq2}{\mbox{sup}}
\left\|\frac{\displaystyle J_{f}({\bf x})^{-1}D^k {f} ({\bf x})}{\displaystyle k!}\right\|^{\frac{1}{k-1}}.
\end{array}$$}
In $\gamma(f,{\bf x})$, the term $D^k f({\bf x})$ is the symmetric tensor whose components are the partial derivatives
of $f$ of order $k$.
Additionally, for convenience, if at some ${\bf x}\in\mathcal{V}(f)$ where $J_{f}({\bf x})$ is not invertible,
then $\alpha(f,{\bf x}):=0$, $\beta(f,{\bf x}):=0$ and $\gamma(f,{\bf x}):=\infty$.
If ${\bf x}\notin\mathcal{V}(f)$ such that $J_{f}({\bf x})$ is not invertible,
then $\alpha(f,{\bf x})$, $\beta(f,{\bf x})$ and $\gamma(f,{\bf x})$ are taken as $\infty$.
Finally, if ${\bf x}$ is an approximate solution of $f$, then $\|{\bf x}-{\bf z}\|\leq2\beta(f,{\bf x})$
where ${\bf z}\in\mathcal{V}(f)$ is the associated solution to ${\bf x}$.

We remark that since this theorem provides a sufficient condition
for a point to be an approximate solution, the set of
{\em certifiable} approximate solutions is generally
much smaller than the set of approximate solutions.
However, for a true approximate solution, a few Newton iterations
usually generate a point that one is able to certify.

Given two approximate solutions ${\bf x}_1$ and ${\bf x}_2$, one often
needs to verify that the corresponding associated solutions ${\bf z}_1$ and~${\bf z}_2$
are distinct.  One way to verify this is by using
the triangle inequality together with
$\|{\bf x}_i-{\bf z}_i\|\leq2\beta(f,{\bf x}_i)$.

\noindent\textbf{Other Nonlinear Systems: }
The above theorem was actually proved with ``polynomial'' replaced by ``analytic.''
However, we present it in this fashion since, in the polynomial case, $\gamma(f,{\bf x})$
is actually defined as a maximum over finitely many terms, since only finitely
many partial derivatives can be nonzero.  In fact, it can be bounded above based
on the coefficients of $f$, the degree of the polynomials in $f$, and $J_f({\bf x})$.
Nonetheless, $\gamma(f,{\bf x})$ can also be bounded above for other nonlinear systems,
in particular, systems of polynomial-exponential equations \cite{hauenstein2011certifying}.
A system is polynomial-exponential if it is polynomial in both the
variables $x_{1},\dots,x_{n}$ and finitely many exponentials of the
form $e^{a x_{i}}$ where $a\in\mathbb{C}$.
Many standard functions such as $\sin(x)$, $\cos(x)$, $\sinh(x)$, and $\cosh(x)$
can be formulated as systems of polynomial-exponential functions since they
are indeed polynomial functions of $e^{a x}$ for~suitable~$a\in\mathbb{C}$.

\noindent\textbf{Real Solutions: }
For a square system $f$ such that $N_f$ defines a real map,
i.e., $N_f({\bf x})$ is real whenever ${\bf x}$ is real, then
Smale's $\alpha$-theory can be extended to provide more information
about real solutions \cite{2010arXiv1011.1091H}.
For potential energy landscapes, when the potential energy function $V({\bf x})$ is real
for real ${\bf x}$, the corresponding Newton iteration is always a real map.
In this case, one can determine the reality of the associated solution ${\bf y}$
from any approximate solution ${\bf x}$.  If ${\bf x}$ is real, then ${\bf y}$ must also be real.
However, if ${\bf x}$ is not real, one can show that ${\bf y}$ is real
by showing that ${\bf x}$ and its real part, namely $({\bf x} + \overline{{\bf x}})/2$
where $\overline{{\bf x}}$ is the conjugate of ${\bf x}$, have the same associated solution, namely ${\bf y}$.
To show that ${\bf y}$ is not real, one simply has to show
that ${\bf x}$ and its conjugate $\overline{{\bf x}}$ have distinct associated
solutions, namely ${\bf y}$ and $\overline{{\bf y}}$, which is shown
using $\|{\bf x}-{\bf y}\|\leq 2\beta(f,{\bf x})$ with the triangle inequality.

We use a recently developed practical implementation of the $\alpha$-theory,
called \textbf{alphaCertified}, for certifying solutions to systems
of equations \cite{2010arXiv1011.1091H,hauenstein2011certifying}.
When using exact rational arithmetic,
the implementation of $\alpha$-theory in \textbf{alphaCertified}
is rigorous and can be taken as a
\textit{mathematical proof of the computed results}.  Hence, this
approach provides an alternative to other analytic or symbolic computations.
The algorithms are also implemented in arbitrary precision floating
point arithmetic in \textbf{alphaCertified}, which provides
certified results up to~round-off~errors.


\noindent\textbf{An Illustrative Example: }
As a demonstration of computing $\alpha(f,x)$, $\beta(f,x)$, and $\gamma(f,x)$
for a single coordinate,
consider the univariate polynomial $f(x) = x^4 - 1$.  In this simple case,
we can actually compute these quantities as a function of a variable $x$
rather than at a specific value.  We will assume $x\neq0$
since $f'(x) = 4x^3$ is zero if and only if $x = 0$ and $f(0)\neq0$.
Clearly, $\beta(f,x) = |x - x^{-3}|/4$.
Now, in the univariate case, the term $D^k f(x)$ in $\gamma(f,x)$
is simply the $k$-th derivative of $f$ at $x$, i.e., $f^{(k)}(x)$.
Since $f$ has degree $4$, we only need to take the maximum over $k = 2,3,4$ to compute $\gamma(f,x)$.
One can easily verify that the maximum is attained at $k = 2$ with
$\gamma(f,x) = 3|x^{-1}|/2$.  Thus, $\alpha(f,x) = 3|1 - x^{-4}|/8$
for any $x\neq0$.

For example, $\alpha(f,2.5) = 0.3654$ so that $x = 2.5$ cannot be certified
as an approximate solution.  In fact, $x = 2.5$ is indeed outside of all of the
quadratic convergence basins.  However, since $\alpha(f,1.1) = 0.11887$,
$x = 1.1$ is certifiably an approximate solution of $f = 0$.  In this case,
we know that the associated solution is $z = 1$ and the following table confirms
the quadratic convergence for small values of $k$. 

{\footnotesize
$$
\begin{array}{c|c|c|c|c|c}
k & 1 & 2 & 3 & 4 & 5\\
\hline
-\log_{10}\left(\|N_{f}^{k}(x)-z\|\right) & 1.89 & 3.62 & 7.06 & 13.94 & 27.70\\
\hline
-\log_{10}\left(\|x-z\|/2^{2^k-1}\right)  & 1.30 & 1.90 & 3.11 & 5.52 & 10.33
\end{array}$$
}

\noindent\textbf{Example With Close Roots: } The $n$-th
Chebyshev polynomial of the first kind is well-known to have $n$
roots between $-1$ and $1$.  These roots, called Chebyshev nodes, are
located at $x_i = \cos\left[(2i-1)\pi/2n\right]$ for $i = 1,\dots,n$.
We can use
this example to demonstrate how small perturbations in a
numerical approximation can change which root Newton's method will converge to.
This chaotic behavior can be avoided by using certification.
In particular, the following table considers selected values where
$f(x) = \cos(50 \cos^{-1}x)$ is the $50$-th Chebyshev polynomial
of the first kind.

{\footnotesize
$$\begin{array}{l|l}
\multicolumn{1}{c|}{x^*} & \multicolumn{1}{c}{\displaystyle\lim_{k\rightarrow\infty} N_f^k(x^*)~~~~} \\
\hline
0.997 & x_2 = \cos(3\pi/100) \\
0.9979 & x_3 = \cos(5\pi/100) \\
0.99799 & x_5 = \cos(9\pi/100) \\
0.997999 & x_6 = \cos(11\pi/100) \\
0.998001 & x_6 = \cos(11\pi/100) \\
0.99801 & x_9 = \cos(17\pi/100) \\
0.9981 & x_1 = \cos(\pi/100) \\
0.998 & x_6 = \cos(11\pi/100)
\end{array}$$}

\noindent\textbf{M\"uller-Brown Surface:}
The M\"uller-Brown surface \cite{muller1979location} is a well-known
model landscape \cite{sunr93,sunr94,wales94a,DoyeW02}. It is defined as
{\footnotesize
\[
V(x,y)= \sum_{i=1}^{4}A_{i}\exp\left(a_{i}(x-x_{i}^{0})^{2}+b_{i}(x-x_{i}^{0})(y-y_{i}^{0})+c_{i}(y-y_{i}^{0})^{2}\right),
\]}
where
{\small
$$\begin{array}{lcllcl}
A&=&(-200, -100, -170, 15), & a&=&(-1, -1, -6.5, 0.7),\\
b&=&(0, 0, 11, 0.6), & c&=&(-10, -10, -6.5, 0.7),\\
x^{0}&=&(1, 0, -0.5, -1), & y^{0}&=&(0, 0.5, 1.5, 1).
\end{array}$$}

Since $\nabla V = [\partial V/\partial x,~\partial V/\partial y]$ involves
polynomials as exponents, we simply add new variables to produce an equivalent
polynomial-exponential form as
{\scriptsize
\[
\begin{array}{l}
f(x,y,z_1,\dots,z_4,w_1,\dots,w_4) = \\
\left[\begin{array}{cl}
\sum_{i=1}^4 A_i w_i (2 a_i (x-x_i^0) + b_i (y - y_i^0)) \\
\sum_{i=1}^4 A_i w_i (b_i (x - x_i^0) + 2 c_i (y - y_i^0))\\
a_i(x-x_i^0)^2 + b_i(x-x_i^0)(y-y_i^0) + c_i(y-y_i^0)^2 - z_i,& i = 1,\dots,4\\
\exp(z_i) - w_i,&i = 1,\dots,4
\end{array}\right].
\end{array}
\]}

Given $(x,y)$, we obtain values of $z_i$ and $w_i$ based on the last
eight functions in $f$ and then try to certify the result.
In particular, the following table presents five numerical approximations
of SPs for $V$ along with an upper bound on the value
of $\alpha(f,\cdot)$ and an approximation of $\beta(f,\cdot)$.
{\footnotesize
$$\begin{array}{c|c|c|l}
 & & \hbox{upper bound} & \multicolumn{1}{c}{\hbox{approximation}}\\
x & y & \hbox{of~}\alpha(f,\cdot) & \multicolumn{1}{c}{\hbox{of~}\beta(f,\cdot)}\\
\hline
-0.5582236346 & 1.441725842 &  0.0140 & ~~~4.84\cdot 10^{-9}\\
0.6234994049 & 0.02803775853 & 0.0460 & ~~~1.94\cdot 10^{-9}\\
0.212486582 & 0.2929883251 &   0.0437 & ~~~3.05\cdot 10^{-9}\\
-0.8220015587 & 0.6243128028 & 0.0006 & ~~~6.94\cdot 10^{-10}\\
-0.050010823 & 0.4666941049 &  0.0068 & ~~~2.89\cdot 10^{-9}
\end{array}$$}
In particular, based on the upper bounds on $\alpha(f,\cdot)$
computed by {\bf alphaCertified}, each
point is indeed an approximate solution.
The bounds on the distance from each
numerical approximation to the corresponding
approximate solution based on $\beta(f,\cdot)$
show that each one must correspond
to a distinct approximate solution.
Thus, we have \textit{proved} that the five numerically approximated SPs are
indeed in the quadratic convergence basin of distinct SPs.

\noindent\textbf{Lennard-Jones Clusters:}
We now consider one of the most studied family of
systems in molecular science, namely atomic clusters
of $N$ atoms bound
by the Lennard-Jones potential \cite{jonesi25}, denoted
LJ$_{N}$.
The pairwise potential between interacting particles is defined as
{\small
$$V_N = 4\epsilon \sum_{i=1}^N\sum_{j=i+1}^N\left[\left(\frac{\sigma}{r_{i,j}}\right)^{12} - \left(\frac{\sigma}{r_{i,j}}\right)^{6} \right],$$}
where $\epsilon$ is the pair well depth, $2^{1/6}\sigma$ is the
equilibrium pair separation, and
{\small
$$r_{i,j} = \sqrt{(x_i - x_j)^2 + (y_i - y_j)^2 + (z_i-z_j)^2}.$$}
For convenience, we take $\epsilon = 1/4$ and $\sigma = 1$.
Since $V_N$ only depends
upon the pairwise distances, the set of SPs is invariant under overall translation
and rotation.  Thus,~we~fix
\begin{equation}\label{eq:FixedCoord}
x_1 = y_1 = z_1 = y_2 = z_2 = z_3 = 0.
\end{equation}
Now, to construct a polynomial system equivalent to $\nabla V_N = {\bf 0}$,
we add variables $R_{i,j}$ with polynomial equations
\begin{equation}\label{eq:Rij}
R_{i,j}\left((x_i - x_j)^2 + (y_i - y_j)^2 + (z_i-z_j)^2\right) = 1.
\end{equation}
That is, $R_{i,j} = 1/r_{i,j}^2$ so that
$V_N = \sum_{i<j} \left(R_{i,j}^6 - R_{i,j}^3\right)$.
For simplicity, we define $R_{i,j} = R_{j,i}$ for $i \neq j$.
Hence, for the SPs, we consider the polynomial system
{\footnotesize
$$\begin{array}{l}
f_N({\bf x},{\bf y},{\bf z},R_{i,j}) = \\
\left[\begin{array}{cl}
\sum_{j \neq i} 6 R_{i,j}^4 \left(2R_{i,j}^3 - 1\right)(x_j - x_i), & i = 2,\dots,N \\
\sum_{j \neq i} 6 R_{i,j}^4 \left(2R_{i,j}^3 - 1\right)(y_j - y_i), & i = 3,\dots,N \\
\sum_{j \neq i} 6 R_{i,j}^4 \left(2R_{i,j}^3 - 1\right)(z_j - z_i), & i = 4,\dots,N \\
R_{i,j}\left((x_i - x_j)^2 + (y_i - y_j)^2 + (z_i-z_j)^2\right) - 1, & i < j
\end{array}\right].
\end{array}$$}
An extensive search for minima and saddle points was carried out
in \cite{2002JChPh.116.3777D} for this model up to $N = 13$
along with a corresponding search for minima and saddles of index one
(transition states) for $N = 14$ in \cite{2005JChPh.122h4105D}.
All of the minima and transition states, available for download at
{\tt http://doye.chem.ox.ac.uk/networks/LJn.html},
were obtained using numerical methods and hence they are numerical approximations.

To certify these solutions, we first translated and rotated each
so that condition \eqref{eq:FixedCoord} holds, and computed $R_{i,j}$
based on \eqref{eq:Rij}. The downloaded points are provided
to 10 decimal places, and many of them were \textit{not} certifiable. We performed two Newton iterations
using 96-bit precision to improve both the precision
and accuracy so that \eqref{eq:FixedCoord} and \eqref{eq:Rij} hold.
Finally, using the resulting points, we employed \textbf{alphaCertified}
to compute upper bounds on $\alpha(f_N,\cdot)$, which we summarize in the following
table.  In particular, this table shows that, for $N = 7,\dots,14$,
each numerical approximation of the minima and transition states does indeed correspond
to a certified approximate solution.
For $N = 14$, the values of
$\left\lceil \log_{10} \gamma(f_{14},\cdot) \right\rceil$ for the minima and transition states
suggest that we should use numerical approaches that approximate the
coordinates of each stationary point to at least $15$ decimal places.  
Hence, certification can also provide insight into convergence
conditions, which hitherto have been chosen based on physical intuition.

Since we performed two Newton
iterations prior to certification, we need to
perform an {\em a posteriori} verification that we still
have distinct solutions.  This was accomplished
using the triangle inequality, as discussed above,
with the maximum value of $\beta(f_N,\cdot)$
and the minimum pairwise distance between the $x$, $y$, $z$ coordinates of the approximations.
To summarize, the following table \textit{proves} that the
numerically approximated SPs are indeed in the
quadratic convergence basin of distinct SPs.

{\footnotesize
$$\begin{array}{c|c|c|c|c|c}
  & & \hbox{maximum} &  & \hbox{maximum} & \hbox{minimum} \\
  & \hbox{number} & \hbox{upper bound} & \hbox{maximum} & \hbox{upper bound} & \hbox{pairwise} \\
N & \hbox{of points} & \hbox{of~}\alpha(f_N,\cdot) & \beta(f_N,\cdot) & \hbox{of~}\gamma(f_N,\cdot) & \hbox{distance} \\
\hline
7 & 16 & 6.82\cdot10^{-19} & 1.03\cdot10^{-28} & 1.01\cdot10^{10} & 0.4354 \\
8 & 50 & 2.03\cdot10^{-19} & 1.28\cdot10^{-28} & 1.60\cdot10^9 & 0.4268 \\
9 & 186 & 3.55\cdot10^{-17} & 1.77\cdot10^{-27} & 5.46\cdot10^{10}  & 0.0559 \\
10 & 699 & 2.86\cdot10^{-14} & 3.94\cdot10^{-24} & 9.87\cdot10^{10} & 0.0600 \\
11 & 2594 & 6.40\cdot10^{-15} & 2.80\cdot10^{-26} & 6.04\cdot10^{11} & 0.0556 \\
12 & 9122 & 1.05\cdot10^{-9} & 1.63\cdot10^{-21} & 4.28\cdot10^{12} & 0.0093 \\
13 & 30265 & 2.48\cdot10^{-12} & 2.84\cdot10^{-23} & 2.16\cdot10^{13} & 0.0081 \\
14 & 91415 & 4.54\cdot10^{-8} & 1.50\cdot10^{-19} & 3.58\cdot10^{14} & 0.0087 \\
\end{array}$$}

\textbf{Conclusion: } Numerical approximate solutions obtained from
standard non-linear optimization methods may lie in the linear convergence basin,
or even in a chaotic region, instead of the desired quadratic region of convergence.
Hence, the numerical approximation may turn out to be a non-solution of the system
when more Newton iterations are performed, which could change the scientific conclusions
drastically. We have demonstrated several examples of such behaviour.
To mitigate such problems, we shown how Smale's $\alpha$-theory
can be used to certify that a numerical approximation is in the quadratic convergence
region of a solution, to determine
if two points correspond to distinct solutions, and to determine if the
corresponding solution is real.
As a practical demonstration of the approach, we have refined and then certified all
the known minima and transition states for the Lennard-Jones potential for up to $14$ atoms.
This is the first certification conducted for a set of physically relevant 
atomic structures that we are aware of,
and it provides quantitative convergence criteria for geometry optimization.
We also observe that for the stationary points of the Lennard-Jones potential, 
the size of the quadratic convergence basin decreases as $N$ increases.
All of these new insights should be applicable throughout molecular science
and studies of soft and condensed matter,
wherever stationary points are considered to analyze structure, dynamics and thermodynamic
properties.

DM was financially supported by the US Department of Energy under contract no. DE-FG02-85ER40237.
JDH would like to thank the US National Science Foundation and Air Force Office of
Scientific Research for their support through DMS-1262428 and FA8650-13-1-7317, respectively. \vspace{-0.75cm}


\begin{thebibliography}{0}%
\makeatletter
\providecommand \@ifxundefined [1]{%
 \@ifx{#1\undefined}
}%
\providecommand \@ifnum [1]{%
 \ifnum #1\expandafter \@firstoftwo
 \else \expandafter \@secondoftwo
 \fi
}%
\providecommand \@ifx [1]{%
 \ifx #1\expandafter \@firstoftwo
 \else \expandafter \@secondoftwo
 \fi
}%
\providecommand \natexlab [1]{#1}%
\providecommand \enquote  [1]{``#1''}%
\providecommand \bibnamefont  [1]{#1}%
\providecommand \bibfnamefont [1]{#1}%
\providecommand \citenamefont [1]{#1}%
\providecommand \href@noop [0]{\@secondoftwo}%
\providecommand \href [0]{\begingroup \@sanitize@url \@href}%
\providecommand \@href[1]{\@@startlink{#1}\@@href}%
\providecommand \@@href[1]{\endgroup#1\@@endlink}%
\providecommand \@sanitize@url [0]{\catcode `\\12\catcode `\$12\catcode
  `\&12\catcode `\#12\catcode `\^12\catcode `\_12\catcode `\%12\relax}%
\providecommand \@@startlink[1]{}%
\providecommand \@@endlink[0]{}%
\providecommand \url  [0]{\begingroup\@sanitize@url \@url }%
\providecommand \@url [1]{\endgroup\@href {#1}{\urlprefix }}%
\providecommand \urlprefix  [0]{URL }%
\providecommand \Eprint [0]{\href }%
\providecommand \doibase [0]{http://dx.doi.org/}%
\providecommand \selectlanguage [0]{\@gobble}%
\providecommand \bibinfo  [0]{\@secondoftwo}%
\providecommand \bibfield  [0]{\@secondoftwo}%
\providecommand \translation [1]{[#1]}%
\providecommand \BibitemOpen [0]{}%
\providecommand \bibitemStop [0]{}%
\providecommand \bibitemNoStop [0]{.\EOS\space}%
\providecommand \EOS [0]{\spacefactor3000\relax}%
\providecommand \BibitemShut  [1]{\csname bibitem#1\endcsname}%
\let\auto@bib@innerbib\@empty
\end{thebibliography}%


\begin{thebibliography}{10}

\bibitem{Wales:04}
D.~J. Wales.
\newblock {\em Energy Landscapes : Applications to Clusters, Biomolecules and
  Glasses (Cambridge Molecular Science)}.
\newblock {Cambridge University Press}, January 2004.

\bibitem{RevModPhys.80.167}
M. Kastner.
\newblock Phase transitions and configuration space topology.
\newblock {\em Rev. Mod. Phys.}, 80(1):167--187, 2008.

\bibitem{mezey81b}
P.~G. Mezey.
\newblock Catchment region partitioning of energy hypersurfaces, i.
\newblock {\em Theo. Chim. Acta}, 58:309, 1981.

\bibitem{mezey87}
P.~G. Mezey.
\newblock {\em Potential Energy Hypersurfaces}.
\newblock Elsevier, Amsterdam, 1987.

\bibitem{wales92}
D.~J. Wales.
\newblock Basins of attraction for stationary-points on a potential-energy
  surface.
\newblock {\em J. Chem. Soc. Faraday Trans.}, 88:653--657, 1992.

\bibitem{wales93d}
D.~J. Wales.
\newblock Locating stationary-points for clusters in cartesian coordinates.
\newblock {\em J. Chem. Soc. Faraday Trans.}, 89:1305--1313, 1993.

\bibitem{Mehta:2009}
D. Mehta.
\newblock {Lattice vs. Continuum: Landau Gauge Fixing and 't Hooft-Polyakov
  Monopoles}.
\newblock {\em Ph.D. Thesis, The Uni. of Adelaide, Australasian Digital Theses
  Program}, 2009.

\bibitem{Mehta:2009zv}
D. Mehta, A. Sternbeck, L. von Smekal, and A.~G. Williams.
\newblock {Lattice Landau Gauge and Algebraic Geometry}.
\newblock {\em PoS}, QCD-TNT09:025, 2009.

\bibitem{Mehta:2011xs}
D. Mehta.
\newblock {Finding All the Stationary Points of a Potential Energy Landscape
  via Numerical Polynomial Homotopy Continuation Method}.
\newblock {\em Phys.Rev.}, E (R) 84:025702, 2011.

\bibitem{Mehta:2011wj}
D. Mehta.
\newblock {Numerical Polynomial Homotopy Continuation Method and String Vacua}.
\newblock {\em Adv.High Energy Phys.}, 2011:263937, 2011.

\bibitem{Kastner:2011zz}
M. Kastner and D. Mehta.
\newblock {Phase Transitions Detached from Stationary Points of the Energy
  Landscape}.
\newblock {\em Phys.Rev.Lett.}, 107:160602, 2011.

\bibitem{Maniatis:2012ex}
M. Maniatis and D. Mehta.
\newblock {Minimizing Higgs Potentials via Numerical Polynomial Homotopy
  Continuation}.
\newblock {\em Eur.Phys.J.Plus}, 127:91, 2012.

\bibitem{Mehta:2012wk}
D. Mehta, Y.-H. He, and J.~D. Hauenstein.
\newblock {Numerical Algebraic Geometry: A New Perspective on String and Gauge
  Theories}.
\newblock {\em JHEP}, 1207:018, 2012.

\bibitem{Hughes:2012hg}
C. Hughes, D. Mehta, and J.-I. Skullerud.
\newblock {Enumerating Gribov copies on the lattice}.
\newblock 2012.

\bibitem{Mehta:2012qr}
D. Mehta, J.~D. Hauenstein, and M. Kastner.
\newblock {Energy landscape analysis of the two-dimensional nearest-neighbor
  $\phi^4$ model}.
\newblock {\em Phys.Rev.}, E85:061103, 2012.

\bibitem{He:2013yk}
Y.-H. He, D. Mehta, M. Niemerg, M. Rummel, and A. Valeanu.
\newblock {Exploring the Potential Energy Landscape Over a Large
  Parameter-Space}.
\newblock 2013.

\bibitem{SW:05}
A.~J. Sommese and C.~W. Wampler.
\newblock {\em The numerical solution of systems of polynomials arising in
  Engineering and Science}.
\newblock World Scientific Publishing Company, 2005.

\bibitem{BCSS}
L.~Blum, F.~Cucker, M.~Shub, and S.~Smale.
\newblock {\em Complexity and real computation}.
\newblock Springer-Verlag, New York, 1998.
\newblock With a foreword by Richard M. Karp.

\bibitem{2010arXiv1011.1091H}
J.~D. Hauenstein and F. Sottile.
\newblock {Algorithm 921: alphaCertified: certifying solutions to polynomial
  systems}.
\newblock {\em ACM TOMS}, 38:28, 2012.

\bibitem{hauenstein2011certifying}
J.~D. Hauenstein and V. Levandovskyy.
\newblock Certifying solutions to square systems of polynomial-exponential
  equations.
\newblock {\em arXiv:1109.4547}, 2011.

\bibitem{muller1979location}
K.~M{\"u}ller and L.~D. Brown.
\newblock Location of saddle points and minimum energy paths by a constrained
  simplex optimization procedure.
\newblock {\em Theoretical Chemistry Accounts: Theory, Computation, and
  Modeling (Theoretica Chimica Acta)}, 53(1):75--93, 1979.

\bibitem{sunr93}
J.-Q. Sun and K.~Ruedenberg.
\newblock Gradient extremals and steepest-descent lines on potential energy
  surfaces.
\newblock {\em J. Chem. Phys.}, 98:9707, 1993.

\bibitem{sunr94}
J.-Q. Sun and K.~Ruedenberg.
\newblock Erratum: Gradient extremals and steepest descent lines on potential
  energy surfaces.
\newblock {\em J. Chem. Phys.}, 100:1779, 1994.

\bibitem{wales94a}
D.~J. Wales.
\newblock Rearrangements of 55-atom lennard-jones and (c-60)(55) clusters.
\newblock {\em J. Chem. Phys.}, 101:3750--3762, 1994.

\bibitem{DoyeW02}
J.~P.~K. Doye and D.~J. Wales.
\newblock Saddle points and dynamics of lennard-jones clusters, solids, and
  supercooled liquids.
\newblock {\em J. Chem. Phys.}, 116:3777--3788, 2002.

\bibitem{jonesi25}
J.~E. Jones and A.~E. Ingham.
\newblock On the calculation of certain crystal potential constants, and on the
  cubic crystal of least potential energy.
\newblock {\em Proc. Roy. Soc. London A}, 107:636--653, 1925.

\bibitem{2002JChPh.116.3777D}
J.~P.~K. {Doye} and D.~J. {Wales}.
\newblock {Saddle points and dynamics of Lennard-Jones clusters, solids, and
  supercooled liquids}.
\newblock {\em Journal of Chem. Phys.}, 116:3777--3788, 2002.

\bibitem{2005JChPh.122h4105D}
J.~P.~K. {Doye} and C.~P. {Massen}.
\newblock {Characterizing the network topology of the energy landscapes of
  atomic clusters}.
\newblock {\em \jcp}, 122(8):084105, February 2005.

\end{thebibliography}

\end{document}